\documentclass[prd,aps,showpacs,nofootinbib,showkeywords,eqsecnum]{revtex4-1}

\usepackage{graphicx,color,amsmath,amsxtra}
\usepackage{epsf}
\usepackage{amssymb}
\usepackage{enumerate}
\usepackage{hhline}
\usepackage{array}
\usepackage{tabularx}
\usepackage[unicode]{hyperref}
\usepackage{graphicx}                
\usepackage{epstopdf}

\begin{document}
\pagestyle{myheadings}

\title{No-go theorem for inflation in Ricci-inverse gravity}
\author{Tuan Q. Do }
\email{tuan.doquoc@phenikaa-uni.edu.vn}
\affiliation{Phenikaa Institute for Advanced Study, Phenikaa University, Hanoi 12116, Vietnam}
\affiliation{Faculty of Basic Sciences, Phenikaa University, Hanoi 12116, Vietnam}
\date{\today} 

\begin{abstract}
In this paper, we study the so-called Ricci-inverse gravity, which is a very novel type of fourth-order gravity proposed recently. In particular, we are able to figure out both isotropically and anisotropically inflating universes to this model. More interestingly, these solutions are shown to be free from a singularity problem. However, stability analysis based on the dynamical system method shows that both isotropic and anisotropic inflation of this model turn out to be unstable against field perturbations. This result implies a no-go theorem for both isotropic and anisotropic inflation in the Ricci-inverse gravity.
\end{abstract}

%


\maketitle
\section{Introduction} \label{intro}
Modern cosmology has experienced a golden age thanks to its recent rapid developments, both in theoretical and observational aspects. Indeed, many interesting results, both for the early time and the late time phases of our universe, have been archived. For the early time phase, the cosmic inflation proposed four decades ago \cite{Starobinsky:1980te,guth} has been regarded as a leading paradigm due to the fact that many its theoretical predictions  have been well confirmed by the recent cosmic microwave background radiations (CMB) observations of the Wilkinson Microwave Anisotropy Probe satellite (WMAP) ~\cite{WMAP} as well as the  Planck one~\cite{Planck}. For the late time phase, the recent observations of accelerated expansion have ultimately changed our understanding on the dynamics of the current universe \cite{cosmic-acceleration,Abbott:2018wog,Scolnic:2017caz}. In fact, the cosmic acceleration  leads us to two theoretical possibilities that:  (i) the modification of Einstein's gravity on the large scales is needed or (ii) the existence of the so-called dark energy assuming the Einstein's gravity is valid for large scales. The first possibility leads to the proposal of alternative (or modified) gravity theories such as the $f(R)$ \cite{DeFelice:2010aj,Nojiri:2010wj,Amendola:2006kh,Appleby:2009uf} and the $f(R,R_{\mu\nu}R^{\mu\nu},R_{\alpha\beta\gamma\delta}R^{\alpha\beta\gamma\delta})$ gravity theories  \cite{Carroll:2004de}. The latter one might address the existence of the cosmological constant $\Lambda$ \cite{Peebles:2002gy} or extra dynamical fields such as the quintessence  field \cite{Caldwell:1997ii} or the phantom field \cite{Caldwell:1999ew}, or the other types \cite{Copeland:2006wr}.

In this paper, we prefer studying a modification of Einstein's gravity. The reason for this is basically twofold. First, this approach does not require the existence of extra fields such as a scalar field, whose origin might  not be easy to figure out. Second,  we note that among the well-known inflationary models the Starobinsky model  \cite{Starobinsky:1980te}, see also Refs. \cite{Whitt:1984pd,Barrow:1983rx,Starobinsky:1987zz,Barrow:1988xh,Maeda:1987xf,Muller:1989rp,Koshelev:2017tvv,Mishra:2019ymr}, involving the $R^2$ correction term has been shown to be one of the most favorable models in the light  of the Planck observation \cite{Planck}.  More interestingly, this model is one of the simplest subclasses of the $f(R)$ theory \cite{DeFelice:2010aj,Nojiri:2010wj} as well as the fourth-order gravity (a.k.a. quadratic gravity) \cite{Schmidt:2006jt}. Hence, having a good alternative gravity model might lead us to a more transparent picture of both early time and late time epochs of our universe \cite{Nojiri:2010wj,Appleby:2009uf}. In addition, unexpected results might appear in alternative gravity theories due to the existence of additional correction terms of fourth-order gravity such as $R^2$ or $R_{\mu\nu}R^{\mu\nu}$. For example, the Hawking's cosmic no-hair conjecture concerning on the isotropy and homogeneity of the late time universe \cite{GH} has been examined in higher curvature and higher derivative gravity \cite{Mijic:1987bq,barrow05,barrow06,Middleton:2010bv,kao09,Toporensky:2006kc,Muller:2017nxg}.  It would be very interesting if we were able to figure out stable anisotropic inflation, which would be counterexamples to the cosmic no-hair conjecture within the framework of modifications of Einstein gravity. Note that some CMB anomalies such as the hemispherical asymmetry and the cold spot, which have been detected recently  by the WMAP and then by the Planck, which cannot be explained within the context of the cosmological principle \cite{Schwarz:2015cma}. Hence, anisotropic inflation might be a possible approach to realize these anomalies. Other anisotropic inflation models, in which the cosmic no-hair conjecture has been shown to be violated, can be seen, e.g. in Refs. \cite{MW,Do:2011zza,SD,Do:2017onf}. If successful, some exotic features of anisotropic inflation  might be imprinted in the CMB, which might be relevant to more sensitive primordial gravitational waves observations operated in the near future \cite{CMB,CMB1}.  

Recently, there has existed a novel gravity model called the Ricci-inverse gravity, which is basically based on the introduction of an {\it anticurvature scalar} $A$, a very new geometrical object \cite{Amendola:2020qho}.  As a result, the anticurvature scalar $A$ is the trace of {\it anticurvature tensor} $A^{\mu\nu}$, which is assumed to be equal to the inverse Ricci tensor, i.e.,  $A^{\mu\nu} =R_{\mu\nu}^{-1}$. This model is a type of fourth-order gravity, similar to models of $R^2$, $R_{\mu\nu}R^{\mu\nu}$,  and  $R_{\mu\nu\alpha\beta}R^{\mu\nu\alpha\beta}$ studied in various papers \cite{Whitt:1984pd,Barrow:1983rx,Starobinsky:1987zz,Barrow:1988xh,Maeda:1987xf,Muller:1989rp,Koshelev:2017tvv,Mishra:2019ymr,Mijic:1987bq,barrow05,barrow06,Middleton:2010bv,kao09,Toporensky:2006kc,Muller:2017nxg}.  Despite the fact that the so-called Ostrogradsky ghost could arise due to the existence of the higher derivatives \cite{Woodard:2015zca}, the fourth-order gravity has played a central role among alternatives to the Einstein gravity \cite{Schmidt:2006jt}. It is worth noting that, the Starobinsky model \cite{Starobinsky:1980te}, one of the simplest fourth-order gravity models, turns out to be free of the Ostrogradsky ghost \cite{Woodard:2015zca}. Additionally, an inflationary solution found in this model has been shown to be highly consistent with the Planck observation ~\cite{Planck}. Hence, it would be interesting if one might construct other fourth-order gravity models, which might exhibit similar properties to the Starobinsky model.  As a result,  the Ricci-inverse gravity model has been shown to admit a no-go theorem claiming that a decelerated and an accelerated expansions cannot exist together in this model. Consequently, this theorem implies that the Ricci-inverse gravity model cannot be a dark energy candidate \cite{Amendola:2020qho}. One  might therefore think of a possibility that the early time inflationary phase of the universe may be suitable for the Ricci-inverse gravity \cite{Amendola:2020qho}. All of these things lead us to investigate whether cosmic inflation appears within the framework of the Ricci-inverse gravity. As a result, we will be able to show  that this model does admit both isotropic and anisotropic inflation, which are really singularity-free. Unfortunately, stability analysis will be performed to indicate that both isotropic and anisotropic inflation are always unstable against field perturbations. This result implies another no-go theorem for the Ricci-inverse gravity that it might be not compatible with the inflationary phase of the universe. Hence, extensions of the Ricci-inverse gravity such as $A\to f(A)$ as proposed in Ref.  \cite{Amendola:2020qho} might be necessary for obtaining (an)isotropic inflation without instabilities.

This paper will be organized as follows: (i) A brief introduction of the present study has been written in the Sec. \ref{intro}. (ii) Basic setup of this model will be presented in Sec. \ref{sec2}. (iii) Simple cosmological solutions to this model will be solved in Sec. \ref{sec3}. (iv) Stability analysis of the obtained solutions will be performed in Sec. \ref{sec4}. (v) Finally, concluding remarks will be written in Sec. \ref{final}.
\section{Basic setup} \label{sec2}
\subsection{Action}
As a result, an action of the so-called Ricci-inverse gravity has been proposed in Ref. \cite{Amendola:2020qho} as follows
\begin{equation} \label{action}
S=\int d^4 x \sqrt{-g}  \left(R+\alpha A -2\Lambda\right),
\end{equation}
where $M_p$ is the reduced Planck mass set to be one for convenience and $\Lambda >0 $ is the pure cosmological constant \cite{barrow06,Toporensky:2006kc,Muller:2017nxg}. In addition, $\alpha$ is a free parameter and $A$ is an anticurvature scalar, which is the trace of anticurvature tensor $A^{\mu\nu}$ assumed to be the inverse Ricci tensor \cite{Amendola:2020qho}
\begin{equation}
A^{\mu\nu} =R_{\mu\nu}^{-1}.
\end{equation}
It is important to note that $A \neq R^{-1}$. As a result, the corresponding Einstein field equation turns out to be \cite{Amendola:2020qho}
\begin{equation} \label{Einstein-field-equation}
R^{\mu\nu}-\frac{1}{2}Rg^{\mu\nu} +\Lambda g^{\mu\nu} = \alpha  A^{\mu\nu} + \frac{1}{2}\alpha A g^{\mu\nu} - \frac{\alpha}{2} \left(2g^{\rho\mu}\nabla_\alpha \nabla_\rho A^{\alpha}_\sigma A^{\nu\sigma} -\nabla^2 A^{\mu}_\sigma A^{\nu\sigma} -g^{\mu\nu}\nabla_\alpha \nabla_\rho A^\alpha_\sigma A^{\rho\sigma}\right).
\end{equation}
where $A^\alpha_\sigma A^{\nu\sigma} =A^{\alpha \tau}g_{\tau\sigma}A^{\sigma \nu} =A^{\alpha\tau}A_\tau^\nu =A^{\alpha\sigma}A_\sigma^\nu =A^\nu_\sigma A^{\alpha \sigma}$. In addition, $\nabla_\mu$ is the covariant derivative.  

Unfortunately, the modified Einstein field equation looks very complicated to derive its explicit non-vanishing components for a given metric. Indeed, the right hand side of Eq. \eqref{Einstein-field-equation} requires a very lengthy calculation, even for the Friedmann-Lemaitre-Robertson-Walker (FLRW) metric \cite{Amendola:2020qho}. Hence, we will not use this tensorial approach but use an effective approach based on the Euler-Lagrange equations. In particular, we will define the Lagrangian of the Ricci-inverse gravity, i.e., 
\begin{equation}
{\cal L} = \sqrt{-g} \left(R+\alpha A -2\Lambda \right),
\end{equation}
 then define the corresponding Euler-Lagrange equations, which are exactly the desired field equations.
\subsection{Field equations} 
In this paper, we propose to consider the ($y-z$ rotational symmetry) Bianchi type I spacetime, which is homogeneous but anisotropic and is described by the following  metric \cite{MW,bianchi},
\begin{equation} \label{metric}
ds^2 =-N^2(t)dt^2 +\exp[2\beta(t)-4\sigma(t)]dx^2 +\exp[2\beta(t)+2\sigma(t)] (dy^2 +dz^2),
\end{equation}
where  $N(t)$ is the lapse function introduced to obtain the following Friedmann equations from its Euler-Lagrange equation \cite{Toporensky:2006kc,Kao:1991zz}. Note that we can set $N= 1$ after deriving its corresponding Friedmann equation \cite{Toporensky:2006kc,Kao:1991zz}. In addition, $\beta(t)$ is an isotropic scale factor and $\sigma(t)$ is regarded as a deviation from isotropy. Hence, $\sigma(t)$ should be much smaller than $\beta(t)$ \cite{MW,Do:2011zza,SD}. Note that the metric shown in Eq. \eqref{metric} is a special case of the Bianchi type I metric in Ref. \cite{barrow06} with $\sigma=\sigma_+$ and $\sigma_- = 0$. The reason is that we would like to figure out analytical solutions for the scale factors of the present model like that we have done in Refs. \cite{MW,Do:2011zza}.

As a result, we are able to define the corresponding non-vanishing components of the Ricci tensor, $R_{\mu\nu}\equiv R^\rho{ }_{\mu\rho\nu}$, to be
\begin{align}
R_{00}&= 3 \left( \frac{\dot N}{N} \dot\beta - \Phi \right), \\
R_{11}&= -\frac{g_{11}}{N^2} \left[ \frac{\dot N}{N} \left(\dot\beta -2\dot\sigma \right)- \Pi \right], \\
R_{22}&=R_{33}= -\frac{g_{33}}{N^2} \left[ \frac{\dot N}{N} \left(\dot\beta + \dot\sigma \right)-\Psi \right],
\end{align}
where we have introduced new variables as
\begin{align}
\Phi &= \ddot\beta +\dot\beta^2 +2\dot\sigma^2,\\
\Pi&=\ddot\beta -2\ddot\sigma+3\dot\beta^2-6\dot\beta\dot\sigma,\\
\Psi&=\ddot\beta +\ddot\sigma+3\dot\beta^2+3\dot\beta\dot\sigma,
\end{align}
for convenience. Note that $\dot\beta \equiv d\beta/dt$, $\ddot\beta \equiv d^2\beta/dt^2$, and $\beta^{(3)} \equiv d^3\beta/dt^3$.
As a result, the corresponding Ricci scalar  $R \equiv g^{\mu\nu}R_{\mu\nu}$ is given by
\begin{equation} \label{def-of-R}
R = -\frac{6}{N^2}\left[  \frac{\dot N}{N} \dot\beta -  \left(\ddot\beta +2\dot\beta^2 +\dot\sigma^2 \right) \right].
\end{equation}
Consequently, the corresponding inverse Ricci scalar $A\equiv g_{\mu\nu}A^{\mu\nu}$ is given by
\begin{align}\label{def-of-A}
A  =& - \frac{N^2}{3} \left( \frac{\dot N}{N} \dot\beta - \Phi \right) ^{-1} -{N^2 } \left[ \frac{\dot N}{N} \left(\dot\beta -2\dot\sigma \right)- \Pi \right]^{-1} -{2 N^2} \left[ \frac{\dot N}{N} \left(\dot\beta + \dot\sigma \right)-\Psi\right]^{-1}.
\end{align}
It is clear that $A\neq R^{-1}$.
Thanks to these useful definitions, we are able to define explicitly the Lagrangian
\begin{equation}
 {\cal L}= \exp[3\beta] N\left(R+\alpha A-2\Lambda \right).
 \end{equation} 
 It appears that this Lagrangian contains three independent variables, $N(t)$, $\beta(t)$, and $\sigma(t)$, along with their time derivatives. Now, we would like to define their corresponding Euler-Lagrange  equations in order to figure out cosmological solutions. First, for the lapse function $N$, we have the following Euler-Lagrange equation defined as
\begin{equation}
\frac{\partial {\cal L}}{\partial N} - \frac{d}{dt}\left( \frac{\partial {\cal L}}{\partial \dot N} \right)=0,
\end{equation}
which will be reduced to
\begin{align} \label{field-equation-1}
 & \alpha \left[\frac{1}{\Phi} - \frac{1}{3\Phi^2}\left(\ddot\beta+3\dot\beta^2 \right) +\frac{2}{3\Phi^3}\dot\beta \left( \beta^{(3)} + 2\dot\beta \ddot\beta +4 \dot\sigma \ddot\sigma \right)\right] +\alpha \left\{ \frac{2}{\Pi} + \frac{2}{\Pi^3}\left(\dot\beta -2\dot\sigma\right) \left[ \beta^{(3)} -2  \sigma^{(3)} +6 \left(\dot\beta - \dot\sigma \right) \ddot\beta - 6 \dot \beta \ddot\sigma \right] \right\}\nonumber\\
&+\alpha \left\{ \frac{4}{\Psi} + \frac{4}{\Psi^3}\left(\dot\beta+\dot\sigma \right)  \left[ \beta^{(3)} +\sigma^{(3)}  +3 \left(2\dot\beta +\dot\sigma \right) \ddot\beta+ 3\dot\beta \ddot\sigma \right] \right\}  + 2\left(3 \dot\beta^2 -3\dot\sigma^2-\Lambda\right)=0,
\end{align}
after setting $N=1$. 
On the other hand, since ${\cal L}$ contains not only $\beta$ and $\dot\beta$ but also $\ddot\beta$, the following Euler-Lagrange equation of $\beta$ turns out to be
\begin{equation}
\frac{\partial {\cal L}}{\partial \beta} - \frac{d}{dt} \left(\frac{\partial {\cal L}}{\partial \dot \beta} \right)+  \frac{d^2}{dt^2}\left( \frac{\partial {\cal L}}{\partial \ddot \beta} \right)=0,
\end{equation}
which will be reduced to
\begin{align}\label{field-equation-2}
 & \alpha \left[ \frac{1}{\Phi} -\frac{1}{3\Phi^2} \left(\ddot\beta +3\dot\beta^2 \right)+\frac{2}{3\Phi^3} \left(\beta^{(4)} +6 \dot\beta \beta^{(3)} +4 \dot\sigma \sigma^{(3)} +2\ddot\beta^2 +4\ddot\sigma^2 +8 \dot\beta^2 \ddot\beta +16 \dot\beta \dot\sigma \ddot\sigma \right)  -\frac{2}{\Phi^4} \left( \beta^{(3)} +2\dot\beta \ddot\beta +4\dot\sigma \ddot\sigma \right)^2 \right] \nonumber\\
& +\alpha \left\{ \frac{6}{\Pi} +\frac{2}{\Pi^3} \left[ \beta^{(4)}-2\sigma^{(4)} +6\dot\beta\beta^{(3)} -6\left(\dot\beta+2\dot\sigma \right)\sigma^{(3)} +6\ddot\beta \left(\ddot\beta-2\ddot\sigma \right) +36\dot\sigma \ddot\beta \left(\dot\beta -\dot\sigma \right) -36\dot\beta\dot\sigma \ddot\sigma \right] \right. \nonumber\\
&\left. -\frac{6}{\Pi^4} \left[ \beta^{(3)} -2\sigma^{(3)} +6\ddot\beta \left(\dot\beta -\dot\sigma \right) -6\dot\beta \ddot\sigma \right]^2\right\} \nonumber\\
& +\alpha \left\{ \frac{12}{\Psi}  + \frac{4}{\Psi^3} \left[\beta^{(4)} +\sigma^{(4)} +6\dot\beta \beta^{(3)} +3\left(\dot\beta -\dot\sigma\right) \sigma^{(3)}  +6\ddot\beta \left(\ddot\beta +\ddot\sigma \right) - 9\dot\sigma \ddot\beta \left(2\dot\beta +\dot\sigma \right) -9\dot\beta \dot\sigma \ddot\sigma \right] \right.\nonumber\\
&\left. -\frac{12}{\Psi^4} \left[ \beta^{(3)} +\sigma^{(3)} + 3 \ddot\beta \left(2\dot\beta+\dot\sigma\right) +3\dot\beta\ddot\sigma \right]^2\right\} +6\left(2\ddot\beta +3\dot\beta^2 +3 \dot\sigma^2 -\Lambda \right)=0.
\end{align}
Finally, the following Euler-Lagrange equation of $\sigma$,
\begin{equation}
\frac{\partial {\cal L}}{\partial \sigma} - \frac{d}{dt} \left(\frac{\partial {\cal L}}{\partial \dot \sigma} \right)+  \frac{d^2}{dt^2}\left( \frac{\partial {\cal L}}{\partial \ddot \sigma} \right)=0,
\end{equation}
leads to 
\begin{align}\label{field-equation-3}
 &\alpha \left[ \frac{4}{3\Phi^2} \left(\ddot\sigma +3\dot\sigma \dot\beta \right) -\frac{8}{3\Phi^3} \dot\sigma \left(\beta^{(3)} +2\dot\beta \ddot\beta +4\dot\sigma \ddot\sigma \right)\right] \nonumber\\
 & -\alpha \left\{ \frac{4}{\Pi^3} \left[ \beta^{(4)} -2\sigma^{(4)} +3 \beta^{(3)} \left(3\dot\beta -2\dot\sigma \right) -12\dot\beta \sigma^{(3)} +6\ddot\beta \left( \ddot\beta -2\ddot\sigma \right) +18 \dot\beta \ddot\beta \left(\dot\beta -\dot\sigma \right) -18\dot\beta^2 \ddot\sigma \right] \right. \nonumber\\
&\left. -\frac{12}{\Pi^4} \left[ \beta^{(3)} -2\sigma^{(3)} +6\ddot\beta \left(\dot\beta-\dot\sigma \right)-6\dot\beta \ddot\sigma \right]^2 \right\} \nonumber\\
&+\alpha \left\{ \frac{4}{\Psi^3} \left[ \beta^{(4)} +\sigma^{(4)} +3\beta^{(3)} \left(3\dot\beta +\dot\sigma \right) +6\dot\beta \sigma^{(3)} +6\ddot\beta \left(\ddot\beta+\ddot\sigma \right) +9\dot\beta \ddot\beta \left(2\dot\beta+\dot\sigma \right) +9\dot\beta^2  \ddot\sigma \right] \right. \nonumber\\
& \left. -\frac{12}{\Psi^4} \left[ \beta^{(3)} +\sigma^{(3)} +3\ddot\beta \left(2\dot\beta +\dot\sigma \right) +3\dot\beta \ddot\sigma \right]^2 \right\} -12\left(\ddot\sigma+3 \dot\beta \dot\sigma \right)=0.
\end{align}
Up to now, we have derived explicit field equations \eqref{field-equation-1}, \eqref{field-equation-2}, and \eqref{field-equation-3} using the effective Euler-Lagrange equation approach. It is apparent that the first one, Eq. \eqref{field-equation-1}, is the Friedmann equation, which is third-order differential equation of $\beta$ and $\sigma$. On the other hand, the other field equations,  Eqs. \eqref{field-equation-2} and \eqref{field-equation-3} are all fourth-order differential equations of $\beta$ and $\sigma$. Hence, solving analytically these field equations is not straightforward task. Fortunately, we are able, thanks to the study done in Ref. \cite{barrow06}, to figure out analytical solutions to these field equation. Detailed cosmological solutions will be shown in the next sections.
\section{Simple exponential solutions} \label{sec3}
\subsection{Singularities}
First, let us briefly present here the singularity issue raised in Ref. \cite{Amendola:2020qho} for expanding universe. If we consider the  Friedmann-Lemaitre-Robertson-Walker (FLRW)  metric of the general following form,
\begin{equation}
ds^2 =-dt^2 +a^2(t) \left(dx^2+dy^2+dz^2 \right),
\end{equation}
where $a(t)$ is the scale factor, we can define the following anticurvature scalar $A$ as
\begin{equation}
A = \frac{2(5\xi+6)}{3H^2 (\xi+1)(\xi+3)},
\end{equation}
where $\xi = H'/H$ with $H' \equiv dH/d\log a = \dot H/H$.
It is clear that $A$ will blow up when either $\xi \to -1$ or $\xi \to -3$, while it will vanish if $\xi= -1.2$. Furthermore, recent observations have claimed that the universe  evolved from the decelerated phase with $\xi \simeq -1.5$ to accelerated phase with $\xi\simeq -0.45$ \cite{Planck,Abbott:2018wog,Scolnic:2017caz,Amendola:2020qho}. It is clear that $-1 \in [-1.5,-0.45]$, resulting the no-go theorem for the Ricci-inverse gravity  in \cite{Amendola:2020qho}.
However, this no-go theorem seems to hold only for expanding universe. For an inflationary phase of universe, it  turns out that $H\simeq \text{constant}$, or equivalently $\xi \simeq 0$, then $A$ is really free from spatial singularities. Indeed, it will become clear for an exponential expansion case shown below. 

 As a result, we will assume the following ansatz for the scale factors as \cite{barrow06} 
\begin{equation} \label{ansatz}
\beta(t) =\zeta t, ~ \sigma(t)=\eta t,
\end{equation}
where $t$ is the cosmic time, while $\zeta$ and $\eta$ are undetermined constants. As a result, the corresponding anticurvature scalar $A$ turns out to be
\begin{equation} \label{def-of-A}
A =\frac{1}{3 \left(\zeta^2+2\eta^2 \right)}+\frac{2}{3\zeta \left(\zeta+\eta \right)} +\frac{1}{3\zeta \left(\zeta -2\eta \right)}.
\end{equation}
  According to Eq. \eqref{def-of-A}, the singularities of $A$ exist at three special points, $\zeta =-\eta$, $\zeta=0$, and $\zeta =2\eta$. However, all these possibilities are not relevant to anisotropic inflationary universes, in which $\eta$ should be much smaller than $\zeta$. Furthermore, for an isotropic universe with a vanishing $\eta$, it turns out that
\begin{equation}
A= \frac{4}{3\zeta^2},
\end{equation}
which is always regular except at a point $\zeta =0$ corresponding the Minkowski spacetime. These results clearly indicate that the no-go theorem claimed in Ref. \cite{Amendola:2020qho} will not be valid for (an)isotropic inflationary universes. In other words, an inflationary universe described by the action \eqref{action} is really free from spatial singularities.
\subsection{Isotropic solutions}
As a result, plugging the ansatz \eqref{ansatz} into the field equation \eqref{field-equation-3} leads to the corresponding algebraic equation of $\zeta$ and $\eta$ such as
\begin{equation} \label{field-equation-4}
\zeta\eta \left[ 9 \left(\zeta^2+2\eta^2 \right)^2 -\alpha \right] =0.
\end{equation}
As a result, a non-trivial solution, $\eta =0$, to Eq. \eqref{field-equation-4} will lead to an isotropic universe. Note that we have ignored the trivial solution, $\zeta =0$. Consequently, both Eqs. \eqref{field-equation-1} and \eqref{field-equation-2} will be reduced to 
\begin{equation} \label{iso-equation-of-zeta}
3\zeta^4 -\Lambda \zeta^2 +\alpha=0.
\end{equation}
Solving this equation will yield two possible solutions,
\begin{align}
\zeta^2_1 &= \frac{1}{6}\left(\Lambda - \sqrt{\Lambda^2 -12\alpha}\right),\\
\zeta^2_2 & = \frac{1}{6}\left(\Lambda + \sqrt{\Lambda^2 -12\alpha}\right).
\end{align}
For a case of non-vanishing $\alpha$, it requires that 
\begin{equation}
\alpha \leq \frac{\Lambda^2}{12}.
\end{equation}
It appears that if $\alpha <0$ then only the solution,
\begin{equation}
\zeta^2_2 = \frac{1}{6}\left(\Lambda + \sqrt{\Lambda^2 -12\alpha}\right) >\frac{\Lambda}{3},
\end{equation}
will be a suitable for describing the early universe. Furthermore, if $|\alpha| \gg \Lambda$, then this solution will reduce to
\begin{equation}
\zeta_2 \simeq \sqrt{\frac{\tilde\Lambda }{3}},
\end{equation}
with $\tilde \Lambda =\sqrt{-\alpha} >0$ as an effective cosmological constant. 

On the other hand, if $0<\alpha \leq \Lambda^2/12$, it will appear that
\begin{align}
&0<\zeta^2_{1}\leq\frac{\Lambda}{6}, \\
&\frac{\Lambda}{6}\leq\zeta^2_{2}<\frac{\Lambda}{3}.
\end{align}
Hence, we can conclude in this case that
\begin{equation}
0<\zeta^2< \frac{\Lambda}{3}.
\end{equation} 
Of course, if $\alpha =0$, we can easily obtain the well-known de Sitter solution with $\zeta^2 =\Lambda/3$. 
\subsection{Anisotropic solutions}
As a result, an anisotropic universe with $\eta \neq 0$ corresponds to a non-trivial solution to Eq. \eqref{field-equation-4}  given by
\begin{equation}
9 \left(\zeta^2+2\eta^2 \right)^2 = \alpha,
\end{equation}
or equivalently
\begin{equation} \label{equation1}
\zeta^2+2\eta^2 = \frac{\kappa}{3},
\end{equation}
where $\kappa =\sqrt{\alpha}>0$. In this case, $\alpha$  have to be positive definite. In other words, the anisotropic solution will not appear for any negative $\alpha$, in contrast to the isotropic solution.

Thanks to the solution \eqref{equation1}, both Eqs. \eqref{field-equation-1} and \eqref{field-equation-2} will reduce to 
\begin{align}
\left(2\kappa -\Lambda \right)\zeta^3 +2\kappa \eta^3 + \Lambda \zeta \eta \left(\zeta+2\eta \right) +\frac{2}{3} \kappa^2 \left(\zeta-2\eta \right) =0.
\end{align}
Furthermore, this equation can be further simplified as
\begin{equation}\label{equation2}
\left(\Lambda-\kappa \right) \left(\zeta+4\eta \right) \zeta \eta+\frac{\kappa}{3}\left(4\kappa -\Lambda \right)\zeta -\kappa^2 \eta =0,
\end{equation}
with the help of the solution \eqref{equation1}.
It appears that we have ended up with two field equations, Eqs. \eqref{equation1} and \eqref{equation2}, for the scale factors $\alpha$ and $\sigma$. 
Since $\sigma$ has been regarded as a deviation from isotropic spacetime, it is expected to be much smaller than $\beta$ during the inflationary phase \cite{MW,Do:2011zza,SD}. Consequently, it turns out that $\eta \ll \zeta$. Note that $\eta$ is not necessarily positive definite. According to Eq. \eqref{equation1}, therefore, we have an approximated solution of $\zeta$,
\begin{equation} \label{zeta}
\zeta \simeq \left( \frac{\kappa}{3}\right)^{1/2}.
\end{equation}
Consequently, Eq. \eqref{equation2} now reduces to
\begin{equation} \label{equation3}
12\left(\kappa-\Lambda \right) \eta^2 +\sqrt{3\kappa}\left(4\kappa -\Lambda \right)\eta -\kappa \left(4\kappa - \Lambda\right) =0.
\end{equation}
As a result, this equation admits two possible solutions,
\begin{equation} \label{equation4}
\eta_{\pm} = \frac{-\sqrt{3\kappa}\left(4\kappa -\Lambda \right)\pm \sqrt{3\left(80\kappa^3 -88\kappa^2\Lambda+17\kappa \Lambda^2 \right)}}{24\left(\kappa-\Lambda \right)}.
\end{equation}

\begin{figure}[hbtp] 
\begin{center}
{\includegraphics[height=60mm]{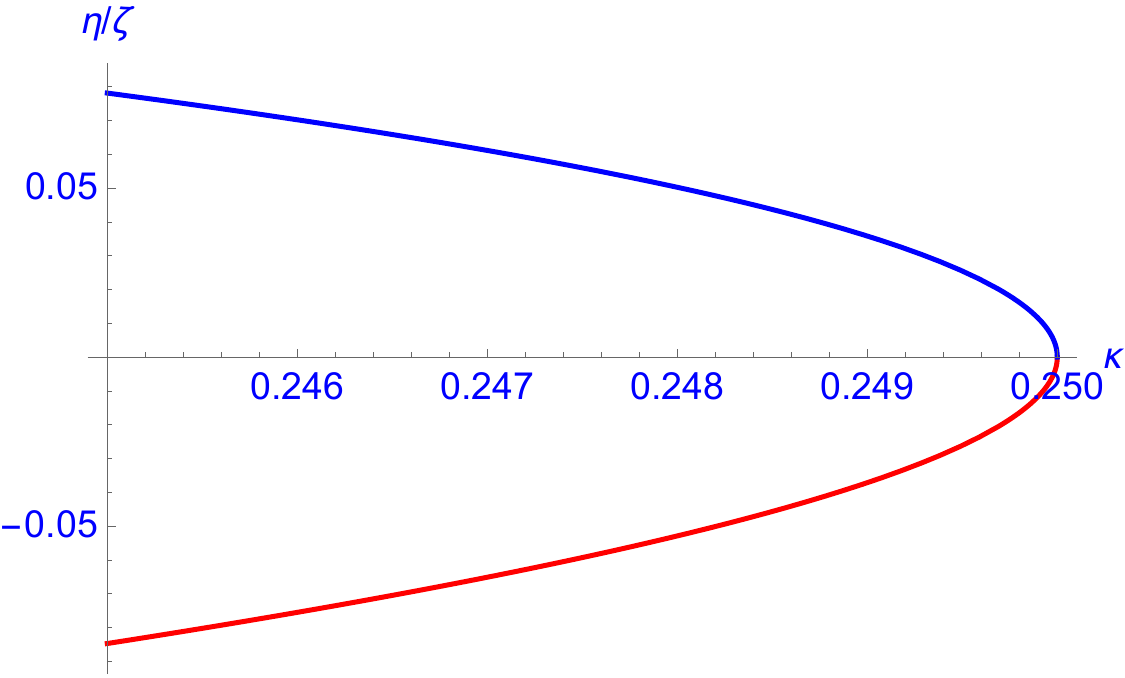}}\\
\caption{A ratio $\eta/\zeta$ as functions of $\kappa$ with $\Lambda=1$. Here the lower red and upper blue  curves correspond to the ratios $\eta_{+}/\zeta$ and $\eta_{-}/\zeta$, respectively.}
\label{fig1}
\end{center}
\end{figure}

It is clear, according to Eq. \eqref{equation3} as well as the solution showin in Eq. \eqref{equation4}, that once $\kappa \to \Lambda /4$ (or equivalently $\alpha \to \Lambda^2/16$) then $\eta \to 0$. This interesting point implies that it is possible to satisfy the following constraint, $\eta\ll \zeta$, when $\kappa$ is close to $ \Lambda /4$. To verify this observation, we will plot below the ratio $\eta/\zeta$ as functions of $\kappa$ with $\Lambda =1$. See the figure \ref{fig1} for details.

In other words, an anisotropic inflation with a small hair (a.k.a. spatial anisotropy) can exist in the Ricci-inverse gravity. Furthermore, if this anisotropic inflation appears as stable and attractor solutions, it would break down the cosmic no-hair conjecture, similar to the anisotropic inflation found in a supergravity-motivated model, in which an unusual coupling between scalar and vector fields, $f^2(\phi)F^{\mu\nu}F_{\mu\nu}$, is introduced \cite{MW,Do:2011zza,SD}. In this case, imprints of anisotropic inflation due to the existence of the anticurvature scalar  $A\equiv g_{\mu\nu}A^{\mu\nu}$ might appear in the CMB and might therefore be detected in the near future by a more sensitive primordial gravitational wave observation \cite{barrow06,CMB,CMB1}. 

Therefore, we will convert the field equations into the corresponding dynamical system of autonomous equations in the next section to examine the stability of the (an)isotropic solutions, following the work done in Ref. \cite{barrow06}. To end this section, we would like to note again that anisotropic inflation with a small spatial anisotropy would happen in this model only for positive $\alpha$ close to $\Lambda^2/16$, while isotropic inflation would exist only when $\alpha <\Lambda^2/12$. This also indicates that only isotropic inflation exists for negative $\alpha$ in this model.
 
\section{Stability analysis} \label{sec4}
\subsection{Dynamical system}
In this section, we would like to investigate the stability of the Bianchi type I inflationary solution within the Ricci-inverse gravity. By doing this, we will convert the field equations, which are  fourth-order differential equations, into the corresponding dynamical system, which is formed by first-order differential equations called the autonomous equations \cite{barrow06}. In particular, we will  introduce dynamical variables as follows \cite{barrow06}
 \begin{align}
 B=\frac{1}{\dot\beta^2},~Q=\frac{\ddot\beta}{\dot\beta^2},~ Q_2 =\frac{\beta^{(3)}}{\dot\beta^3}, ~ \Omega_\Lambda =\frac{\Lambda}{3\dot\beta^2},~ \Sigma =\frac{\dot\sigma}{\dot\beta}, ~\Sigma_1 =\frac{\ddot\sigma}{\dot\beta^2}, ~\Sigma_2 =\frac{\sigma^{(3)}}{\dot\beta^3}, 
 \end{align}
 here the Hubble constant is given by $H=\dot\beta$. 
 As a result, a set of autonomous equations of dynamical variables can be defined to be
 \begin{align}
  \label{Dyn-1}
 B' &= -2QB,\\
  \label{Dyn-2}
 \Omega_\Lambda' &= -2Q\Omega_\Lambda,\\
  \label{Dyn-3}
 Q' &=Q_2 -2Q^2,\\
 \label{Dyn-6}
 Q_2 '&= \frac{\beta^{(4)}}{\dot\beta^4}-3Q Q_2,\\
  \label{Dyn-4}
 \Sigma' &= \Sigma_1 - Q\Sigma ,\\
  \label{Dyn-5}
 \Sigma_1'&= \Sigma_2 -2Q \Sigma_1,\\
  \label{Dyn-7}
 \Sigma_2'&= \frac{\sigma^{(4)}}{\dot\beta^4}-3Q \Sigma_2,
 \end{align}
 where $' \equiv d/d\tau$ with $\tau =\int \dot\beta dt$ being the dynamical time variable. It is noted that the terms ${\beta^{(4)}}/{\dot\beta^4}$ and ${\sigma^{(4)}}/{\dot\beta^4}$ in  two equations \eqref{Dyn-6} and \eqref{Dyn-7} can be figured out from the field equations \eqref{field-equation-2} and \eqref{field-equation-3}, which can be rewritten respectively in terms of the dynamical variables as follows
 \begin{align} \label{beta-sigma-1}
 &\alpha\left[\frac{B}{\Phi}-\frac{1}{3\Phi^2} \left(Q+3\right)+\frac{2}{3B\Phi^3} \left(\frac{\beta^{(4)}}{\dot\beta^4}+6Q_2 +4\Sigma \Sigma_2+2Q^2 +4\Sigma_1^2 +8Q+16\Sigma \Sigma_1 \right) -\frac{2}{B^2 \Phi^4} \left(Q_2 +2Q+4\Sigma \Sigma_1 \right)^2 \right]\nonumber\\
 &+\alpha \left\{ \frac{6B}{\Pi}+\frac{2}{B\Pi^3} \left[\frac{\beta^{(4)}}{\dot\beta^4}-2\frac{\sigma^{(4)}}{\dot\beta^4} +6Q_2 -6 \Sigma_2\left(1+2\Sigma\right) +6Q\left(Q-2\Sigma_1 \right)+36\Sigma Q \left(1-\Sigma\right)-36\Sigma \Sigma_1 \right] \right. \nonumber\\
& \left. -\frac{6}{B^2 \Pi^4} \left[Q_2 -2\Sigma_2 +6Q\left(1-\Sigma \right)-6\Sigma_1 \right]^2 \right\} \nonumber\\
 &+\alpha \left\{ \frac{12B}{\Psi}+\frac{4}{B\Psi^3} \left[\frac{\beta^{(4)}}{\dot\beta^4}+\frac{\sigma^{(4)}}{\dot\beta^4} +6Q_2 +3 \Sigma_2\left(1-\Sigma\right) +6Q\left(Q+\Sigma_1 \right)-9\Sigma Q \left(2+\Sigma\right)-9\Sigma \Sigma_1 \right] \right. \nonumber\\
& \left. -\frac{12}{B^2 \Psi^4} \left[Q_2 +\Sigma_2 +3Q\left(2+\Sigma \right)+3\Sigma_1 \right]^2 \right\} +6\left(2Q+3\Sigma^2 -3\Omega_\Lambda +3 \right)=0,
 \end{align}
 \begin{align} \label{beta-sigma-2}
& \alpha \left[\frac{4}{3\Phi^2} \left(\Sigma_1 +3\Sigma \right) -\frac{8}{3B\Phi^3} \Sigma \left(Q_2 +2Q+4\Sigma\Sigma_1 \right)\right]\nonumber\\
& -\alpha \left\{ \frac{4}{B\Pi^3} \left[ \frac{\beta^{(4)}}{\dot\beta^4}-2\frac{\sigma^{(4)}}{\dot\beta^4} +3Q_2 \left(3-2\Sigma \right) -12\Sigma_2 +6Q \left(Q-2\Sigma_1 \right)+18Q\left(1-\Sigma \right)-18\Sigma_1 \right] \right. \nonumber\\
& \left. -\frac{12}{B^2 \Pi^4} \left[ Q_2 -2\Sigma_2 +6Q \left(1-\Sigma\right) -6\Sigma_1 \right]^2 \right\}\nonumber\\
& -\alpha \left\{ \frac{4}{B\Psi^3} \left[ \frac{\beta^{(4)}}{\dot\beta^4}+\frac{\sigma^{(4)}}{\dot\beta^4} +3Q_2 \left(3+\Sigma \right) +6\Sigma_2 +6Q \left(Q+\Sigma_1 \right)+9Q\left(2+\Sigma \right)+9\Sigma_1 \right] \right. \nonumber\\
& \left. -\frac{12}{B^2 \Psi^4} \left[ Q_2 +\Sigma_2 +3Q \left(2+\Sigma\right) +3\Sigma_1 \right]^2 \right\} -12\left(\Sigma_1 +3\Sigma \right)=0,
\end{align}
where the corresponding $\Phi$, $\Pi$, and $\Psi$ are given by
 \begin{align}
 \Phi &= \frac{1}{B} \left(Q+2\Sigma^2 +1 \right),\\
 \Pi&= \frac{1}{B} \left(Q -2\Sigma_1-6\Sigma +3\right),\\
 \Psi &=  \frac{1}{B} \left(Q +\Sigma_1+3\Sigma +3\right).
 \end{align} 
 It is also noted that the dynamical variables should obey the constraint equation, which is nothing but the Friedmann equation \eqref{field-equation-1},
 \begin{align} \label{beta-sigma-0}
& \alpha \left[\frac{B}{\Phi} -\frac{1}{3\Phi^2} \left(Q+3\right) +\frac{2}{3B\Phi^3} \left(Q_2 +2Q+4\Sigma\Sigma_1 \right)\right]+\alpha \left\{ \frac{2B}{\Pi} +\frac{2}{B\Pi^3} \left(1-2\Sigma\right) \left[ Q_2 -2\Sigma_2 +6 Q \left(1-\Sigma \right) -6\Sigma_1 \right] \right\}\nonumber\\
 & +\alpha \left\{ \frac{4B}{\Psi}+\frac{4}{B\Psi^3} \left(1+\Sigma\right) \left[Q_2 +\Sigma_2 +3Q\left(2+\Sigma \right)+3\Sigma_1 \right]\right\} -2 \left(3\Sigma^2 +3\Omega_\Lambda -3 \right)=0.
 \end{align}
 \subsection{Fixed points}
As a result, the fixed points to the dynamical system of the autonomous equations \eqref{Dyn-1}, \eqref{Dyn-2}, \eqref{Dyn-3},  \eqref{Dyn-6}, \eqref{Dyn-4}, \eqref{Dyn-5}, and \eqref{Dyn-7} are solutions of the following equations, $B'=\Omega_\Lambda' =Q'=Q_2'=\Sigma'=\Sigma_1'  =\Sigma_2'=0$. It appears that the isotropic fixed points correspond to 
\begin{align}
Q=Q_2=\Sigma=\Sigma_1=\Sigma_2 =0
\end{align}
along with the following equation
\begin{equation} \label{iso-fixed-point}
\alpha B^2 -3\Omega_\Lambda +3 =0,
\end{equation}
which can be reduced to
\begin{equation}
3\dot\beta^4 -\Lambda \dot\beta^2 +\alpha =0.
\end{equation}
More interestingly, this equation is identical to Eq. \eqref{iso-equation-of-zeta} with $\dot\beta = \zeta$. This means that the isotropic exponential solutions found in the previous section are indeed equivalent to these isotropic fixed points. Note again that we have shown $\alpha <\Lambda^2/12$ is the constraint for the existence of isotropic inflation. 

For anisotropic fixed points with $\Sigma \neq 0$, it turns out that
\begin{align}
Q=Q_2=\Sigma_1=\Sigma_2 =0
\end{align}
along with the following equations
\begin{align} \label{anisotropic-fixed-point-1}
9\left(2\Sigma^2 +1 \right)^2 &=\alpha B^2,\\
\label{anisotropic-fixed-point-2}
6\kappa B \Sigma^3 +18\Omega_\Lambda \Sigma^2 -\left(4\kappa^2 B^2 -9\Omega_\Lambda\right)\Sigma +2\kappa^2 B^2 +6\kappa B -9\Omega_\Lambda& =0,
\end{align}
where $\kappa =\sqrt{\alpha}> 0$. Here, the first equation is due to the equation \eqref{beta-sigma-2}, while the last equation is derived from both equations \eqref{beta-sigma-1} and \eqref{beta-sigma-0} with the help of the first equation. Similar to the isotropic fixed point, these equations are identical to Eqs. \eqref{equation1} and \eqref{equation2} with $\dot\beta =\zeta$ and $\dot\sigma =\eta$. 
Note that both ${\beta^{(4)}}/{\dot\beta^4}$ and ${\sigma^{(4)}}/{\dot\beta^4}$ vanish for all fixed points due to $Q=0$. It is important to note that an anisotropic inflation should have a small anisotropy, i.e., $|\Sigma| \ll 1$. Consequently, it appears, according to Eqs. \eqref{anisotropic-fixed-point-1} and \eqref{anisotropic-fixed-point-2}, that 
\begin{align} \label{appro}
\alpha B^2 \simeq 9, ~\Omega_\Lambda \simeq 4.
\end{align}

All these results imply that the anisotropic exponential solutions found in the previous section are equivalent to these anisotropic fixed points. Hence, the stability of the fixed points and exponential solutions share the same properties. It is important to note that due to the equation \eqref{anisotropic-fixed-point-1}  the anisotropic fixed points will not appear for any negative $\alpha$, in contrast to the isotropic fixed points. 
\subsection{Stability of isotropic fixed points}
Now, we would like to examine the stability of the fixed points, which are equivalent to the exponential solutions found in the previous section. First, we will consider the isotropic fixed points by perturbing the autonomous equations around them as follows
\begin{align}
\delta B' &=-2B \delta Q,\\
\delta \Omega_\Lambda'&=-2\Omega_\Lambda \delta Q,\\
\delta Q' &= \delta Q_2,\\
\delta Q_2' &= \delta \left(\frac{\beta^{(4)}}{\dot\beta^4}\right),\\
\label{iso-pert-1}
\delta \Sigma' & =\delta \Sigma_1,\\
\label{iso-pert-2}
\delta \Sigma_1'&=\delta \Sigma_2,\\
\label{iso-pert-3}
\delta \Sigma_2' &= \delta \left(\frac{\sigma^{(4)}}{\dot\beta^4}\right).
\end{align}
where $\delta \left({\beta^{(4)}}/{\dot\beta^4}\right)$ and $\delta \left({\sigma^{(4)}}/{\dot\beta^4}\right)$ will be figured out from the following perturbed equations derived from Eqs. \eqref{beta-sigma-1} and \eqref{beta-sigma-2}, 
\begin{align} \label{pert-1}
&\alpha \left\{ \frac{1}{\Phi} \delta B -\frac{B}{\Phi^2}\delta \Phi -\frac{1}{3\Phi^2}\delta Q +\frac{2}{\Phi^3}\delta \Phi  +\frac{2}{3B \Phi^3} \left[ \delta  \left(\frac{\beta^{(4)}}{\dot\beta^4}\right) +6\delta Q_2 +8\delta Q \right] \right\} \nonumber\\
& +\alpha \left\{ \frac{6}{\Pi}\delta B -\frac{6B}{\Pi^2} \delta \Pi +\frac{2}{B \Pi^3} \left[ \delta  \left(\frac{\beta^{(4)}}{\dot\beta^4}\right) - 2\delta  \left(\frac{\sigma^{(4)}}{\dot\beta^4}\right) +6\delta Q_2 -6\delta \Sigma_2 \right] \right\} \nonumber\\
&+\alpha\left\{ \frac{12}{\Psi}\delta B -\frac{12B}{\Psi^2} \delta \Psi +\frac{4}{B \Psi^3} \left[ \delta  \left(\frac{\beta^{(4)}}{\dot\beta^4}\right) +\delta  \left(\frac{\sigma^{(4)}}{\dot\beta^4}\right) +6\delta Q_2 +3\delta \Sigma_2 \right] \right\} + 6 \left(2\delta Q -3\delta \Omega_\Lambda\right)=0,
\end{align}
\begin{align} \label{pert-2}
& \frac{4\alpha}{3\Phi^2} \left(\delta\Sigma_1 +3\delta \Sigma \right) - \frac{4\alpha}{B \Pi^3} \left[ \delta  \left(\frac{\beta^{(4)}}{\dot\beta^4}\right) - 2\delta  \left(\frac{\sigma^{(4)}}{\dot\beta^4}\right) +9\delta Q_2 +18\delta Q -12\delta \Sigma_2 -18 \delta \Sigma_1 \right] \nonumber\\
&- \frac{4\alpha}{B \Psi^3} \left[ \delta  \left(\frac{\beta^{(4)}}{\dot\beta^4}\right) +\delta  \left(\frac{\sigma^{(4)}}{\dot\beta^4}\right) +9\delta Q_2 +18\delta Q +6\delta \Sigma_2 +9 \delta \Sigma_1 \right] -12 \left(\delta \Sigma_1 +3\delta \Sigma \right)=0,
\end{align}
along with the perturbed  Friedmann equation given by
 \begin{align}\label{pert-Fried-1}
& \alpha \left[ \frac{1}{\Phi}\delta B -\frac{B}{\Phi^2}\delta\Phi -\frac{1}{3\Phi^2} \delta Q +\frac{2}{\Phi^3} \delta \Phi+\frac{2}{3B \Phi^3} \left(\delta Q_2 +2\delta Q \right) \right]+\alpha \left[\frac{2}{\Pi}\delta B-\frac{2B}{\Pi^2}\delta \Pi +\frac{2}{B\Pi^3}  \left(\delta Q_2 -2\delta \Sigma_2 +6\delta Q-6\delta \Sigma_1 \right)\right] \nonumber\\
&+\alpha \left[\frac{4}{\Psi} \delta B-\frac{4B}{\Psi^2}\delta \Psi +\frac{4}{B\Psi^3} \left(\delta Q_2 +\delta \Sigma_2 +6\delta Q +3\delta \Sigma_1 \right)\right]-6\delta \Omega_\Lambda =0.
 \end{align}
It is noted that
\begin{align}
\delta \Phi &= -\frac{1}{B^2}\delta B +\frac{1}{B}\delta Q,\\
\delta \Pi &= -\frac{3}{B^2}\delta B +\frac{1}{B} \left(\delta Q-2\delta \Sigma_1 -6\delta \Sigma \right),\\
\delta \Psi &= -\frac{3}{B^2}\delta B +\frac{1}{B} \left(\delta Q+\delta \Sigma_1 +3 \delta \Sigma \right),
\end{align}
as well as 
\begin{equation}
\Phi =\frac{1}{B}, ~\Pi = \Psi =\frac{3}{B}.
\end{equation}

As a result, we can obtain $\delta \Omega_\Lambda$ from Eq. \eqref{pert-Fried-1}  as
\begin{equation}
\delta \Omega_\Lambda = \frac{2\alpha}{27} \left(9B\delta B +6B^2 \delta Q +2B^2\delta Q_2 \right).
\end{equation}
Thanks to this solution, we are able to obtain the following results from Eqs. \eqref{pert-1} and \eqref{pert-2} 
\begin{align}
\delta \left(\frac{\beta^{(4)}}{\dot\beta^4} \right)=&  -\frac{9}{2}\left(\frac{3}{\alpha B^2}-1 \right)\delta Q -3\delta Q_2,\\
\delta \left(\frac{\sigma^{(4)}}{\dot\beta^4} \right)=& -9 \left(\frac{3}{\alpha B^2}-5 \right)\delta Q +12\delta Q_2 +27\left(\frac{9}{\alpha B^2}-1\right)\delta \Sigma + 9\left(\frac{9}{\alpha B^2}-2\right)\delta \Sigma_1 -6\delta \Sigma_2 .
\end{align}
Taking exponential perturbations such as 
\begin{equation}
\delta B, ~\delta \Omega_\Lambda, ~\delta Q, ~\delta Q_2, ~\delta\Sigma, ~\delta \Sigma_1, ~\delta\Sigma_2 \propto \exp\left[\mu \tau \right],
\end{equation}
we are able to obtain the following equation of $\mu$,
 \begin{equation}
 \mu^2 \left(\mu+3\right) \left(\alpha B^2 \mu^2 + 3\alpha B^2 \mu +9 \alpha B^2 -81 \right) \left(2\alpha B^2 \mu^2 +6\alpha B^2 \mu -9\alpha B^2 +27 \right)=0,
 \end{equation}
 where the solution \eqref{iso-fixed-point} has been used to simplify this equation.
 As a result, the corresponding values of $\mu$ are solved to be
\begin{equation}
\mu_{1,2} =0 , ~ \mu_{3} =-3, ~\mu_{4,5} =-\frac{3}{2} \left[1 \pm \frac{\sqrt{3\left(12\alpha B^2 -\alpha^2 B^4 \right)}}{\alpha B^2}\right], ~\mu_{6,7}= -\frac{3}{2} \left[1 \pm \frac{\sqrt{-3\left(2\alpha B^2 -\alpha^2 B^4 \right)}}{\alpha B^2} \right].
\end{equation}

\begin{figure}[hbtp] 
\begin{center}
{\includegraphics[height=70mm]{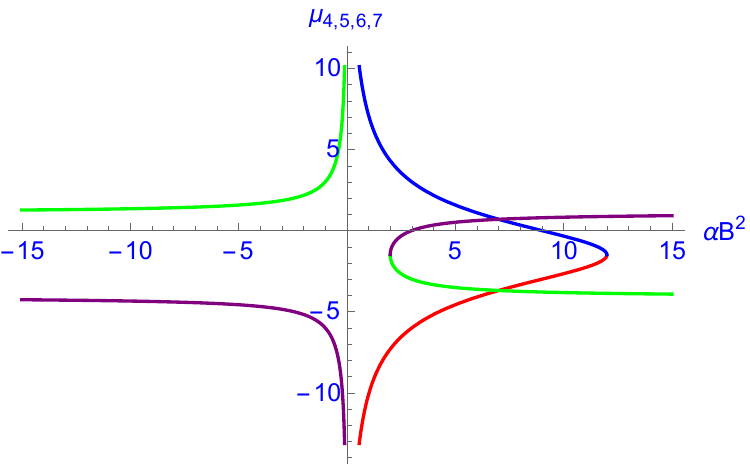}}\\
\caption{Eigenvalues $\mu_4$ (red), $\mu _5$ (blue), $\mu_6$ (green), and $\mu_7$ (purple) as functions of $\alpha B^2$.}
\label{fig2}
\end{center}
\end{figure}

It is straightforward to see that if $\alpha <0$ then the isotropic fixed points are always unstable since $\mu_6 >0$. Now, we consider the case $\alpha >0$. First, we will focus on the range $2\leq \alpha B^2 \leq 12$, in which $\mu_{4,5,6,7}$ are all real. According to the figure \ref{fig2}, it turns out that while $\mu_4$ and $\mu_6$ are always negative  there is always at least one positive $\mu$ among two eigenvalues $\mu_{5}$ and $\mu_7$ in the range $2\leq \alpha B^2 \leq 12$. Hence, the isotropic fixed points will also be unstable in this range of $\alpha B^2$. Then, we will plot $\mu_{4,5}$ in a range $0< \alpha B^2 <2 $ and plot $\mu_{6,7}$ in a range $2<\alpha B^2 \leq 15$ to see whether unstable modes exist. As a result, there is always one unstable mode in these ranges. Hence, we now can conclude that the isotropic fixed points of the Ricci-inverse gravity always turn out to be unstable against field perturbations, in contrast to the quadratic gravity studied in literature \cite{barrow05,barrow06,Middleton:2010bv,kao09,Toporensky:2006kc,Muller:2017nxg}.
\subsection{Stability of anisotropic fixed points}
Before going to investigate the stability of anisotropic fixed points in details, we would like to note that the anisotropic fixed points will not exist for any negative value of $\alpha$. From now on, therefore, we will only consider the positive $\alpha$ case. Note again that $|\Sigma| \ll 1$ for is required for an anisotropic inflation \cite{MW,Do:2011zza,SD}. Therefore, Eq. \eqref{iso-pert-1} should be modified as
\begin{equation}
\delta \Sigma' =\delta \Sigma_1 -\Sigma \delta Q.
\end{equation}
In addition, Eqs. \eqref{pert-1} and \eqref{pert-2} should also be modified for non-vanishing $\Sigma$ as
\begin{align} \label{pert-3}
&\alpha \left\{ \frac{1}{\Phi} \delta B -\frac{B}{\Phi^2}\delta \Phi -\frac{1}{3\Phi^2}\delta Q +\frac{2}{\Phi^3}\delta \Phi  +\frac{2}{3B \Phi^3} \left[ \delta  \left(\frac{\beta^{(4)}}{\dot\beta^4}\right) +6\delta Q_2+4\Sigma \delta \Sigma_2 +8\delta Q +16\Sigma \delta \Sigma_1 \right] \right\} \nonumber\\
& +\alpha \left\{ \frac{6}{\Pi}\delta B -\frac{6B}{\Pi^2} \delta \Pi +\frac{2}{B \Pi^3} \left[ \delta  \left(\frac{\beta^{(4)}}{\dot\beta^4}\right) - 2\delta  \left(\frac{\sigma^{(4)}}{\dot\beta^4}\right) +6\delta Q_2 -6(1+2 \Sigma) \delta \Sigma_2 +36\Sigma(1 -\Sigma) \delta Q -36\Sigma \delta \Sigma_1 \right] \right\} \nonumber\\
&+\alpha\left\{ \frac{12}{\Psi}\delta B -\frac{12B}{\Psi^2} \delta \Psi +\frac{4}{B \Psi^3} \left[ \delta  \left(\frac{\beta^{(4)}}{\dot\beta^4}\right) +\delta  \left(\frac{\sigma^{(4)}}{\dot\beta^4}\right) +6\delta Q_2 +3(1 -\Sigma) \delta \Sigma_2-9\Sigma(2+\Sigma)\delta Q-9\Sigma\delta \Sigma_1 \right] \right\} \nonumber\\
&+ 6 \left(2\delta Q +6\Sigma \delta \Sigma-3\delta \Omega_\Lambda\right)=0,
\end{align}
\begin{align} \label{pert-4}
&\alpha\left[ -\frac{8}{\Phi^3}\Sigma\delta \Phi+\frac{4}{3\Phi^2} \left(\delta\Sigma_1 +3\delta \Sigma \right) -\frac{8}{3B\Phi^3} \Sigma \left(\delta Q_2 +2\delta Q +4\Sigma \delta \Sigma_1 \right) \right] \nonumber\\
&- \frac{4\alpha}{B \Pi^3} \left[ \delta  \left(\frac{\beta^{(4)}}{\dot\beta^4}\right) - 2\delta  \left(\frac{\sigma^{(4)}}{\dot\beta^4}\right) +3(3-2\Sigma)\delta Q_2 +18(1-\Sigma)\delta Q -12\delta \Sigma_2 -18 \delta \Sigma_1 \right] \nonumber\\
&- \frac{4\alpha}{B \Psi^3} \left[ \delta  \left(\frac{\beta^{(4)}}{\dot\beta^4}\right) +\delta  \left(\frac{\sigma^{(4)}}{\dot\beta^4}\right) +3(3+\Sigma)\delta Q_2 +9(2+\Sigma)\delta Q +6\delta \Sigma_2 +9 \delta \Sigma_1 \right] -12 \left(\delta \Sigma_1 +3\delta \Sigma \right)=0,
\end{align}
along with 
\begin{align}
\delta \Phi &= -\frac{1}{B^2}\left(2\Sigma^2+1\right)\delta B +\frac{1}{B}\left(\delta Q+4\Sigma \delta \Sigma \right),\\
\delta \Pi &= \frac{3}{B^2}\left(2\Sigma-1 \right)\delta B +\frac{1}{B} \left(\delta Q-2\delta \Sigma_1 -6\delta \Sigma \right),\\
\delta \Psi &= -\frac{3}{B^2} \left(\Sigma+1 \right)\delta B +\frac{1}{B} \left(\delta Q+\delta \Sigma_1 +3 \delta \Sigma \right),
\end{align}
as well as 
\begin{equation}
\Phi =\frac{1}{B}\left(2\Sigma^2+1\right), ~\Pi  =-\frac{3}{B}\left(2\Sigma-1 \right), ~\Psi =\frac{3}{B}\left(\Sigma+1 \right).
\end{equation}
 In this case, the perturbed equation of the Friedmann equation \eqref{beta-sigma-0} turns out to be
  \begin{align}
& \alpha \left[ \frac{1}{\Phi}\delta B -\frac{B}{\Phi^2}\delta\Phi -\frac{1}{3\Phi^2} \delta Q +\frac{2}{\Phi^3} \delta \Phi+\frac{2}{3B \Phi^3} \left(\delta Q_2 +2\delta Q +4\Sigma \delta \Sigma_1\right) \right]\nonumber\\
&+\alpha \left\{\frac{2}{\Pi}\delta B-\frac{2B}{\Pi^2}\delta \Pi +\frac{2}{B\Pi^3} \left(1-2\Sigma \right) \left[\delta Q_2 -2\delta \Sigma_2 +6(1-\Sigma)\delta Q-6\delta \Sigma_1 \right] \right\} \nonumber\\
&+\alpha \left\{\frac{4}{\Psi} \delta B-\frac{4B}{\Psi^2}\delta \Psi +\frac{4}{B\Psi^3} (1+\Sigma)\left[\delta Q_2 +\delta \Sigma_2 +3(2+\Sigma)\delta Q +3\delta \Sigma_1 \right]\right\}-6\left(2\Sigma \delta \Sigma+\delta \Omega_\Lambda \right)=0,
 \end{align}
which can be solved to give
\begin{equation}
\delta \Omega_\Lambda \simeq \frac{2\alpha}{27} \left(9B\delta B +6B^2 \delta Q +2B^2\delta Q_2 \right) +\frac{2\Sigma}{9}\left[ 3\left(\alpha B^2 -3\right)\delta \Sigma +2\alpha B^2 \delta \Sigma_1 \right]. 
\end{equation}
Here, the constraint $|\Sigma|\ll 1$ has been used to simplify this solution.  As a result, we are able to figure out the following relations for anisotropic fixed points with $|\Sigma|\ll 1$,
\begin{align}
\delta \left(\frac{\beta^{(4)}}{\dot\beta^4} \right) \simeq &  -\frac{9}{2} \left(\frac{3}{\alpha B^2} -1\right)\delta Q -3 \delta Q_2 - {9} \left(\frac{9}{\alpha B^2} -1\right)\Sigma \delta \Sigma +\frac{3}{2}\Sigma \delta \Sigma_1 -3\Sigma \delta \Sigma_2,\\
\delta \left(\frac{\sigma^{(4)}}{\dot\beta^4} \right)\simeq &  -\frac{54}{B}\Sigma \delta B- 9\left(\frac{3}{\alpha B^2}-5 \right)\delta Q + 12 \delta Q_2  + 27 \left(\frac{9}{\alpha B^2} -1 \right)\delta \Sigma +9 \left( \frac{9}{\alpha B^2} -2 \right)\delta \Sigma_1  -6 \delta \Sigma_2.
\end{align}
As a result, besides two trivial eigenvalues,
\begin{equation}
\mu_{1,2} =0,
\end{equation}
there are five other non-trivial eigenvalues determined from the following equation,
\begin{equation} \label{eigen-equation}
f(\mu)\equiv a_5 \mu^5+a_4\mu^4+a_3 \mu^3 +a_2\mu^2 +a_1 \mu +a_0=0,
\end{equation}
with
\begin{equation}
a_5=1, ~a_4=9, ~a_3= 12 \left(3\Sigma+2\right),~a_2=  9 \left(12\Sigma +1 \right),~a_1= 9 \left(36\Sigma^2 -7\Sigma -3 \right), ~a_0= -162\Sigma^2.
\end{equation}
Note that the approximations shown in Eq. \eqref{appro} for an anisotropic inflation have been used to define the above eigenvalue equation of $\mu$. Using the simple method in Ref. \cite{Do:2011zza}, we are able to conclude that Eq. \eqref{eigen-equation} always admits at least one positive root $\mu>0$ without solving it explicitly.  Indeed, it is clear that $f(\mu=0)=a_0 <0$ and $f(\mu \gg 1) \sim a_5 \mu^5 >0$. Therefore, the curve $f(\mu)$ will cross the positive horizontal  $\mu$-axis at least one time at $\mu=\mu_\ast$. And, this intersection point $\mu=\mu_\ast$, which is positive definite, is exactly a root to the equation, $f(\mu)=0$. This result indicates that the anisotropic fixed points with a small anisotropy ($|\Sigma|\ll 1$) turn out to be unstable against field perturbations, consistent with the Bianchi type I inflation found in the quadratic gravity \cite{barrow06} as well as the prediction of the cosmic no-hair conjecture. It should be noted that the instabilities, found in this section for both isotropic and anisotropic solutions, are purely classical and therefore could not be related to the  Ostrogradsky ghost \cite{Woodard:2015zca}. Other  classical instabilities of the Bianchi type I solutions of higher order gravity models, e.g., the Einsteinian cubic gravity \cite{Bueno:2016xff,Erices:2019mkd}, can been seen in Ref. \cite{Pookkillath:2020iqq}.
\section{Conclusions}\label{final}
We have investigated a novel Ricci-inverse gravity proposed in Ref. \cite{Amendola:2020qho} recently, in which a very novel geometrical object $A$ called the anticurvature scalar is introduced. Basically, $A$ is defined in terms of $A^{\mu\nu}$, the anticurvature tensor assumed to be the inverse of Ricci tensor, i.e., $A^{\mu\nu}=R^{-1}_{\mu\nu}$, as $A\equiv g_{\mu\nu}A^{\mu\nu}$. As a result, we have derived the corresponding field equations of this model using the effective Euler-Lagrange equations for the Bianchi type I metric. Then, we have figured out both isotropically and anisotropically inflating solutions to this model. More interestingly, we have shown that these inflationary solutions make the anticurvature scalar  $A$ singularity-free, in contrast to the no-go theorem in \cite{Amendola:2020qho}, which seems to be valid only for the accelerating universe in the late time.  
Stability analysis based on the dynamical system method has been performed to show that the both isotropic and anisotropic inflation turn out to be unstable against field perturbations. This result implies a no-go theorem for both isotropic and anisotropic inflation in the Ricci-inverse gravity. Hence, extensions of the Ricci-inverse gravity, e.g., $A \to f(A)$ proposed in Ref. \cite{Amendola:2020qho}, might be necessary in order to resolve this instability issue. Details of this consideration will be presented in a sequel to this paper. We hope that our study would shed more light on the cosmological implications of the Ricci-inverse gravity, which is a very promising alternative gravity model deserved to investigate more in the near future.
\begin{acknowledgments}
The author would like to thank the referee very much for useful comments and suggestions. The author would also like to thank Dr. Leonardo Giani very much for his correspondence. This study is supported by the Vietnam National Foundation for Science and Technology Development (NAFOSTED) under grant number 103.01-2020.15. 
\end{acknowledgments}

\end{document}